\documentclass[prd,superscriptaddress,onecolumn,showkeys]{revtex4}
\usepackage{eurosym}
\usepackage{amsfonts}
\usepackage{array}
\usepackage{amsthm}
\usepackage{bm}
\usepackage{palatino}
\usepackage{mathpazo}
\usepackage{amssymb}
\usepackage{eurosym}
\usepackage{amsmath}
\usepackage{epsfig}
\usepackage{graphics}
\usepackage{changes}
\usepackage{color}
\usepackage{graphicx}
\usepackage[colorlinks=true,
            linkcolor=blue,
           urlcolor=black,
           citecolor=black]{hyperref}

\def\be{\begin{equation}}
\def\ee{\end{equation}}
\def\beq{\begin{eqnarray}}
\def\eeq{\end{eqnarray}}

\def\bes{\begin{eqnarray}}
\def\ees{\end{eqnarray}}

\begin{document}

\title{\textbf{Gravity Effects on Hawking Radiation from Charged Black Strings in Rastall Theory}}

\author{Riasat Ali}
\email{riasatyasin@gmail.com}
\affiliation{Department of Mathematics, GC
University Faisalabad Layyah Campus, Layyah-31200, Pakistan}

\author{Rimsha Babar}
\email{rimsha.babar10@gmail.com}
\affiliation{Division of Science and Technology, University of Education, Township, Lahore-54590, Pakistan}

\author{Muhammad Asgher}
\email{m.asgher145@gmail.com}
\affiliation{Department of Mathematics, The Islamia
University of Bahawalpur, Bahawalpur-63100, Pakistan}

\author{Syed Asif Ali Shah}
\email{asifalishah695@gmail.com}
\affiliation{Department of Mathematics and Statistics, The University of Lahore 1-Km Raiwind Road,
Sultan Town Lahore 54000, Pakistan}

\begin{abstract}
The Rastall theory of gravity is the generalized form of the Einstein theory which
describes the conservation law of energy and momentum tensor.
In our work, we compute the charged black strings solution in the background of
Rastall theory by applying the Newman-Janis approach.
After computing the charged black strings solution in the background of Rastall
theory, we study the thermodynamical property (i.e., Hawking temperature) for the
charged black strings. Furthermore, we investigate the graphical representation
of Hawking temperature via event horizon to check the stability conditions of charged
black strings under the influence of Rastall theory. Moreover, we examine the modified
Hawking temperature for charged  black strings in Rastall theory by taking into account
the quantum gravity effects. We also discuss the physical state of charged black strings
under the effects of quantum gravity and spin  parameter (appears due to Rastall theory
in charged black strings solution).
\end{abstract}
\keywords{Black strings; Rastall theory; Newman-Janis algorithm; Hawking temperature.}

\date{\today}
\maketitle

\section{Introduction}

General relativity (GR) theory of Einstein, which is assumed to be the most interesting and simplest gravity theory,
obeys the conservation law of energy-momentum tensor.
Although, since its establishment researchers are looking for different gravity theories and several modified gravity theories
have been developed. In this campaign, Rastall \cite{1,2} introduced a very interesting potential modification of
GR theory, which does not obey the standard conservation law of energy-momentum tensor (i.e., $T^{uv}_{;u}=0$).
However, a non-minimal coupling of matter field via space-time geometry can be introduced in the form
\begin{equation}
T^{v}_{u;v}=\lambda R_{,u},\label{zz}
\end{equation}
here $\lambda$ represents the coupling parameter and describes the deviation from GR.
The spherically symmetric static charged as well as uncharged black hole (BH)
metric in the context of perfect fluid surrounded by Rastall gravity theory have been
analyzed \cite{3}. Additionally, some interest has been committed to provide
the static spherically symmetric solutions of the gravitational field equations in the
background of Rastall gravity which incorporates the BH, wormholes and neutron star solutions \cite{3a,3b}.
The Reissner-Nordstr\"{o}m BH metric solution with cosmological constant in Rastall gravity theory
has been studied \cite{4}.

Spallucci and Smailagic \cite{5} have analyzed the regular BH solution in the context of Rastall
gravity theory. They conclude that a regular BH solution exists with exotic matter and have no
singularity in General Relativity. The BH solutions (in the background of perfect fluid matter of rotating BHs) in the Rastall
theory have been analyzed \cite{6, 7}. The spherically symmetric
and static regular BH metric in the generalized Rastall gravity theory have
been investigated \cite{8}. Moreover, the electromagnetic neutral BHs solution and their
general properties have also been analyzed.

The theory of Rastall gravity is a generalized gravity theory and also studied the coupling between geometry and matter.
According to Visser, the Rastall theory of gravity is equivalent to Einstein gravity \cite{8a}
but the Darabi and his colleagues \cite{9b} conclusion is different from Visser's idea.
They proposed that Rastall gravity is not equivalent to Einstein gravity.
The rotating BH solution by utilizing Demia\'{n}ski-Newman-Janis
algorithm to the electrically charged BH surrounded by quintessence parameter in
Rastall gravity theory have been analyzed \cite{9}. Furthermore, the BH mass and thermodynamical
properties (Hawking temperature, heat capacity and electromagnetic potential)
from the horizon equation have also been examined.

Moradpour et al. have analyzed the conformally flat BH solutions in the Rastall gravity
as well as non-singular BH solutions in the background of modified Rastall gravity \cite{10}.
The Hawking radiation depends on BH geometry and for different types of particles, we arrive at the same result.
Yale \cite{20} have studied the Hawking temperature for every type of particle fermions, scalars and bosons spin-$1$,
by utilizing the tunneling method.
The Hawking temperature for symmetric BHs can be derived \cite{20} from the following formula
\begin{equation}
T_{H} = \frac{\acute{f}(r_+)}{4\pi}.\label{a11}
\end{equation}

In order to calculate the Hawking temperature, the Hawking radiation phenomenon for
different BHs have been investigated \cite{11}-\cite{Sakalli:2015nza}. Moreover, they have also studied the Hawking temperature for various types of BHs by taking into account the quantum gravity effects.
By considering the generalized uncertainty principle (GUP) effects, it is feasible to study the quantum corrected thermodynamical properties of BH \cite{20a}. The GUP offers high-energy remedies to BH thermodynamics, which guides to the possibility of a minimal length in quantum theory of gravity. The modified fundamental commutation relation can be describes as $[x_{\mu}, p_{\mu}]=i\hbar\delta_{\mu\nu}[1+\alpha p^2]$ \cite{20b} .

The expression of GUP can be defined as
\begin{equation}
\Delta x\Delta p \geq \frac{\hbar}{2}\left[1+\alpha(\Delta p)^2\right],
\end{equation}
where $\alpha=\frac{\alpha_0}{M_p^2}$, $\alpha_0$ denotes the dimensionless parameter and $\alpha_0<10^{5}$, the ${M_p^2}$ gives the Plank mass.
The $x_{\mu}$ and $p_{\mu}$ denotes the modified position and momentum operators, respectively, which can be given as
$x_{\mu}=x_{0\mu}, p_{\mu}=p_{0\mu}\left(1+\alpha p^2_{0\mu}\right),$
where $p_{0\mu}$ and $x_{0\mu}$ are standard momentum and position operators,
respectively, which satisfy the standard commutation relation $[x_{0\mu}, p_{0\mu}]=i\hbar\delta_{\mu\nu}$.
We choose the values of $\alpha$ according to the condition, which
satisfies the condition of GUP relation \cite{20c}. For corrected Hawking temperature, we have choose
the only first order terms of $\alpha$ in our calculation.
The idea of GUP has been utilized for various BHs in literature \cite{16}-\cite{19}.

The main aim of this article is to study the charged black strings solution in the background of Rastall
theory and to compare our results with previous literature.
This paper is arranged in the following way: In Sec. \textbf{II},
we derive a charged black strings solution in the context of Rastall theory and also investigate
the Hawking temperature for the charged black strings. Section \textbf{III} provides the graphical explanation of
Hawking temperature via event horizon and states the stability condition of charged black strings under
the Rastall theory effects. Section \textbf{IV} analyze the modified Hawking temperature for charged
black strings in the Rastall theory.
Section \textbf{V} discusses the effects of quantum gravity and Rastall parameter on charged black strings
with the help of graphical interpretation.
Finally, Sec. \textbf{VI} consists of summary and conclusions.

\section{Charged Black Strings Solution in the Context of Rastall Theory}
By applying the Demia\'{n}ski-Newman-Janis algorithm, the spin or rotation parameter $a$ can be derived into a spherically symmetric solution which provides an extension for Newman-Janis algorithm.
The Rastall gravity depends upon the Rastall theory that established the conservation law of
energy-momentum to the accompanying framework as described in Eq. (\ref{zz}).
According to this theory, the modified Einstein field equation can be defined as \cite{jj}
\begin{equation}
G_{uv}+\kappa\lambda g_{uv}=\kappa \tilde{T}_{uv},
\end{equation}
where $G_N$ stands for gravitational constant for the Newton gravity, $\kappa=8\pi G_N$
represents the gravitational constant of the Rastall gravity.

Here, we derive a metric for charged static and stationary black strings in the background of Rastall
theory by considering the Newman-Janis algorithm. Moreover, we investigate the Hawking temperature
for the corresponding BH metric. For this purpose, we consider the charged static black strings solution \cite{21}
\begin{equation}
ds^{2}=-F(r)dt^2+\frac{1}{F(r)}dr^2+r^2 d\theta^2+r^2 \beta^2dy^2,\label{sol}
\end{equation}
where
\begin{equation}
F(r)=\beta^2 r^2-\frac{b}{\beta r}+\frac{c^{2}}{\beta^{2} r^{2}},\nonumber
\end{equation}
and
\begin{equation}
\beta^2 =-\frac{\bigwedge}{3},~~~b=4MG,~~~c^{2}=4q^{2}G,\nonumber
\end{equation}
here the parameters $M$ and $q$ shows the ADM mass per unit length and black string charge,
respectively. Moreover, $\bigwedge$ denotes the cosmological constant as well as $b$ and $c$
represents the arbitrary parameters.

After putting $\beta^2 r^2-\frac{b}{\beta r}+\frac{c^{2}}{\beta^{2} r^{2}}=0$, we can
evaluate the event horizon in the given form \cite{22}
\begin{equation}
r_{+}=\frac{S^{\frac{1}{2}}b^{\frac{1}{3}}+2^{\frac{1}{2}}[S^{2}
-4p^{2}-S]^{\frac{1}{4}}}{2\beta},
\end{equation}
where
\begin{eqnarray}
S&=&\left[0.5+0.5\left(1-\frac{256p^{6}}{27}\right)^{\frac{1}{2}}\right]^{\frac{1}{3}}+
\left[0.5-0.5\left(1-\frac{256p^{6}}{27}\right)^{\frac{1}{2}}\right]^{\frac{1}{3}},\nonumber\\
p^{2}&=&\frac{1}{b^{\frac{4}{3}}}c^{2}.\nonumber
\end{eqnarray}
In order to analyze the charged black strings solution in the context of Rastall theory.
Firstly, we consider a transformation for black strings metric Eq. (\ref{sol}) from
coordinates $(t, r, \theta, y)$ to $(u, r, \theta, y)$ as
\begin{eqnarray}
du=dt-\frac{dr}{F(r)},\label{A}
\end{eqnarray}
under the given transformation the Eq. (\ref{sol}) can be defined as
\begin{equation}
ds^{2}=-F(r)du^2-2dudr+r^2 d\theta^2+r^2 \beta^2dy^2.
\end{equation}
The components of the inverse metric can be given as
\begin{equation}
g^{ur}=g^{ru}=-1,~~g^{rr}=F,~~g^{\theta\theta}=\frac{1}{r^2},~~g^{yy}=\frac{1}{r^2\beta^2}.
\end{equation}

The inverse metric in the frame of null tetrad can be expressed as
\begin{eqnarray}
g^{\mu\nu}=-l^\nu n^\mu-l^\mu n^\nu+m^\mu \bar{m}^{\nu}+m^\nu \bar{m}^{\mu}.
\end{eqnarray}
The corresponding elements for null tetrad can be defined in the form
\begin{eqnarray}
l^{\mu}&=&\delta_{r}^{\mu},~~~n^{\mu}=\delta_{u}^{\mu}-\frac{1}{2} F \delta_{r}^{\mu},\nonumber\\
m^{\mu}&=&\frac{1}{\sqrt{2}r} \delta_{\theta}^{\mu}+\frac{i}{\sqrt{2}r\beta}\delta_{y}^{\mu},\nonumber\\
\bar{m}^{\mu}&=&\frac{1}{\sqrt{2}r} \delta_{\theta}^{\mu}-\frac{i}{\sqrt{2}r\beta}\delta_{y}^{\mu},\nonumber
\end{eqnarray}
At any point in the black string metric, the relations between the null tetrad and the null vectors becomes
\begin{equation}
l_{\mu}l^{\mu}=n_{\mu}n^{\mu}=m_{\mu}m^{\mu}=l_{\mu}m^{\mu}=m_{\mu}m^{\mu}=0,\nonumber
\end{equation}
and
\begin{equation}
l_{\mu}n^{\mu}=-m_{\mu}\bar{m}^{\mu}=1.\nonumber
\end{equation}
In the $(u, r)$ plane, the coordinate transformation can be defined as
\begin{eqnarray}
u&\rightarrow &u-ia\cos\theta,\nonumber\\
r&\rightarrow &r+ia\cos\theta,\nonumber
\end{eqnarray}
%where $\lambda$ is already defined %as in Eq. (1).
Moreover, we analyze the following transformations
\begin{equation}
F(r)\rightarrow f(r, a, \theta),
\end{equation}
and
\begin{equation}
r^2+a^2 \cos^2\theta=\Sigma^2.
\end{equation}
In the $(u, r)$ plan the null vectors get the form
\begin{eqnarray}
l^{\mu}&=&\delta_{r}^{\mu},~~~n^{\mu}=\delta_{u}^{\mu}-\frac{1}{2}
f\delta_{r}^{\mu},\nonumber\\
m^{\mu}&=&\frac{1}{\sqrt{2}r}\left(\delta_{\theta}^{\mu}+ia \beta(\delta_{u}^{\mu}
-\delta_{r}^{\mu})+\frac{i}{\beta}\delta_{y}^{\mu}\right),\nonumber\\
\bar{m}^{\mu}&=&\frac{1}{\sqrt{2}r}\left(\delta_{\theta}^{\mu}-ia \beta(\delta_{u}^{\mu}
-\delta_{r}^{\mu})-\frac{i}{\beta}\delta_{y}^{\mu}\right).\nonumber
\end{eqnarray}
According to the null tetrad definition, the non-zero components of inverse metric
$g^{\mu r}$ in the $(u, r, \theta, y)$ coordinates can be derived as
\begin{eqnarray}
g^{uu}&=&\frac{a^2\beta^2}{\sum^2},~~~g^{ur}=g^{ru}=-1-
\frac{a^2\beta^2}{\sum^2},
~~~g^{rr}=f(r, \theta)+\frac{a^2\beta^2}{\sum^2},~~~
\nonumber\\
g^{yy}&=&\frac{1}{\sum^2\beta^2},~~~g^{uy}=g^{yu}=\frac{a}{\sum^2},~~~
g^{ry}=g^{yr}=-\frac{a}{\sum^2},~~~g^{\theta\theta}=\frac{1}{\sum^2},\nonumber
\end{eqnarray}
here
\begin{equation}
f(r, \theta)=\frac{\beta^2r^4-\frac{4Mr}{\beta}+\frac{4q^2}{\beta^2}}{\Sigma^2}.\label{TT}
\end{equation}
Furthermore, we analyze a coordinate transformation from $(u, r, \theta, y)$
to $(t, r, \theta, y)$ coordinates in the given form
\begin{equation}
du=dt+\Lambda(r)dr,~~~dy=dy+h(r)dr\label{lam},
\end{equation}
here \begin{eqnarray}
\Lambda(r)&=&\frac{r^2+a^2}{r^2F +a^2},\nonumber\\
 h(r)&=&-\frac{a}{r^2F +a^2}.\nonumber
\end{eqnarray}
%we get
%\begin{eqnarray}
%ds^{2}&=&-F(r)du^2-2dudr-2a\beta^2(1-F)dud\theta+2a\beta^2drd\theta+\Sigma^2dy^2\nonumber\\
%&+&\beta^2(\Sigma^2-a^2\beta^2(\Phi-2))d\theta^2.
%\end{eqnarray}
%
%\begin{eqnarray}
%ds^{2}&=&-F(r)dt^2-2a\beta^2(1-F)dt d\theta-[F(\lambda(r))^2-2\lambda(r)+\beta^2h(r)\nonumber\\
%&&(\Sigma^2-a^2\beta^2(F-2))]dr^2+\Sigma^2dy^2+\beta^2[\Sigma^2-a^2\beta^2(F-2)]d\theta^2.
%\end{eqnarray}
Finally, we compute the black strings metric in the background of Rastall theory under $(t, r, \theta, y)$ coordinates in the
following form
%Finally, we compute the black strings %metric in the background of Rastall %theory under
%$(t, r, \theta, y)$ coordinates in %the following form
\begin{eqnarray}
ds^{2}&=&-\left(\frac{\beta^2r^4-\frac{4Mr}{\beta}+\frac{4q^2}{\beta^2}}{\Sigma^2}\right)
dt^2-2a\beta^2\left(1-\frac{\beta^2r^4-\frac{4Mr}{\beta}+\frac{4q^2}{\beta^2}}{\Sigma^2}
\right)dt dy+\frac{\Sigma^2}{\Delta_{r}}dr^2\nonumber\\
&+&\Sigma^2 d\theta^2+\frac{a^2\left[\Sigma^4+a^2(-4q^2+(4Mr-r^4\beta^3+2\beta \Sigma^2)
\beta)\right]}{\Sigma^2}dy^2,\label{ww}
\end{eqnarray}
where
\begin{equation*}
\Delta_{r}=\beta^2r^4-\frac{4Mr}{\beta}+\frac{4q^2}{\beta^2}.
\end{equation*}
%\begin{equation}
%T_{H} = \frac{\acute{f}(r_+)}{4\pi}.
%\end{equation}
The generalized formula for the Hawking temperature has been commonly computed in
the previous literature. By using Eq. (\ref{a11}) the Hawking temperature can be evaluated in the following expression
\begin{equation}
T_{H}=\frac{2\beta^4r^5_{+}+4M\beta r^2_{+}-8r_{+}q^2+a^2(4\beta^4r^3_{+}-4M\beta)}{4\pi \beta^2(r^2_+ +a^2)^2}.\label{12}
\end{equation}
The temperature $T_{H}$ depends on cosmological constant $\bigwedge$ (i. e., $\beta=-\bigwedge/3$),
spin parameter $a$, black string mass $M$ and black
string charge $q$. It is worth mentioning here that for $a=0$, we recover the Hawking temperature
for charged black strings \cite{21}, %but changed some order of $r_+$ and %$\beta$ due to transformation
which is independent of the spin parameter.
\begin{equation}
T_{H}=\frac{1}{4\pi}\left[2\beta^2r_{+}+\frac{4M}{\beta r^2_{+}}-\frac{8q^2}{\beta^2r^3_{+}}\right].
\end{equation}

\section{Graphical Analysis}
This section analyzes the graphical explanation of $T_{H}$
w.r.t horizon $r_{+}$. We observe the physical
importance of the graphs under the influence of spin parameter
and study the stability analysis of corresponding charged black strings.
According to Hawking's phenomenon when the temperature increases and more
radiations emit then radius of BH reduces. This physical phenomenon depicts the BH stability.

In \textbf{Fig. 1}: (i) represents the behavior of $T_{H}$ for fixed $M=100$, $\beta=-0.001$,
$a=9$ and varying values of BH charge $q$. It is to be noted that the temperature $T_{H}$
slowly decreases with the increasing values of $r_{+}$ in the range $0\leq r_{+}\leq8$.
This behavior shows the stability of BH.

In (ii), one can observe the behavior of $T_{H}$ for fixed $M=200$, $\beta=-0.0005$,
$q=0.1$ and varying values of spin parameter $a$. Here, we can see that the $T_{H}$ exponentially decreases as $r_{+}$ increases.
Moreover, it can be also seen that in the range $3.1<r_{+}< 3.3$, the temperature remains same for the various values of $a$.
The physical behavior of $T_{H}$ in the range $0\leq r_{+}\leq5$ guarantee the stable condition of BH.

\begin{center}
\includegraphics[width=8.2cm]{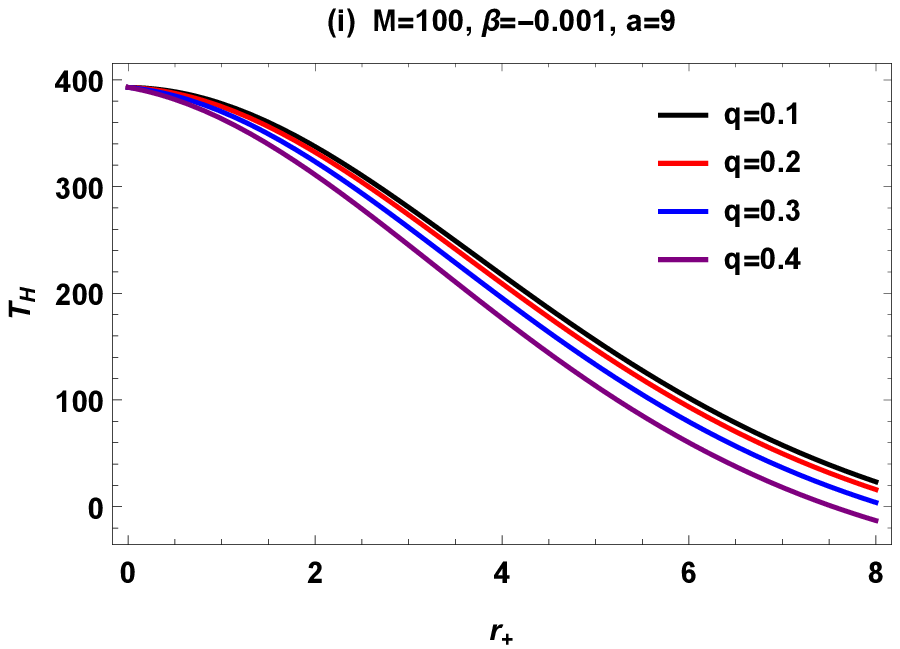}\includegraphics[width=8.2cm]{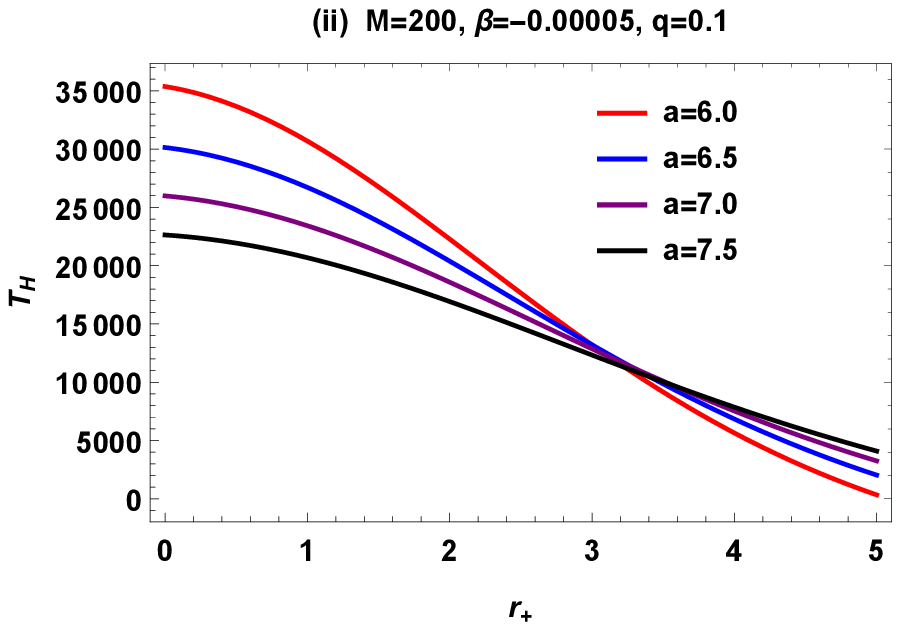}\\
{Figure 1: Hawking temperature $T_{H}$ versus event horizon $r_{+}$.}
\end{center}

\section{Corrected Temperature for Charged Black Strings in Rastall Theory}
This section analyzes the quantum gravity effects on Hawking temperature
of charged black strings in the Rastall theory for massive vector particles.
To do so, we write the Eq. (\ref{ww}) in the following form
\begin{eqnarray}
ds^{2}&=&-Adt^{2}+Bdr^{2}+Cd\theta^{2}
+D dy^{2}+2Edt dy,\label{aa}
\end{eqnarray}
where
\begin{eqnarray}
A&=&\left(\frac{\beta^2r^4-\frac{4Mr}{\beta}+\frac{4q^2}{\beta^2}}{\Sigma^2}\right),
~~D=\frac{a^2\left[\Sigma^4+a^2(-4q^2+(4Mr-r^4\beta^3+2\beta \Sigma^2)
\beta)\right]}{\Sigma^2},\nonumber\\
B&=&\frac{\Sigma^2}{\Delta_{r}},~~~~~~~~~C=\Sigma^2,~~~~~~~~~~~E=-a\beta^2\left(1-\frac{\beta^2r^4
-\frac{4Mr}{\beta}+\frac{4q^2}{\beta^2}}{\Sigma^2}\right).\nonumber
\end{eqnarray}
The equation of wave motion is defined as \cite{19}
\begin{eqnarray}
&&\partial_{\mu}(\sqrt{-g}\chi^{\nu\mu})+\sqrt{-g}\frac{m^2}{\hbar^2}
\chi^{\nu}+\sqrt{-g}\frac{i}{\hbar}A_{\mu}\chi^{\nu\mu}+\sqrt{-g}\frac{i}
{\hbar}eF^{\nu\mu}\chi_{\mu}+\alpha\hbar^{2}\partial_{0}\partial_{0}
\partial_{0}(\sqrt{-g}g^{00}\chi^{0\nu})\nonumber\\
&&-\alpha \hbar^{2}\partial_{i}\partial_{i}\partial_{i}(\sqrt{-g}g^{ii}\chi^{i\nu})=0,\label{xx}
\end{eqnarray}
here $g$ gives the determinant of coefficient matrix, $\chi^{\nu\mu}$ represents the anti-symmetric
tensor and $m$ is the particle mass, since
\begin{eqnarray}
\chi_{\nu\mu}&=&(1-\alpha{\hbar^2\partial_{\nu}^2})\partial_{\nu}\chi_{\mu}-
(1-\alpha{\hbar^2\partial_{\mu}^2})\partial_{\mu}\chi_{\nu}+
(1-\alpha{\hbar^2\partial_{\nu}^2})\frac{i}{\hbar}eA_{\nu}\chi_{\mu}
-(1-\alpha{\hbar^2}\partial_{\nu}^2)\frac{i}{\hbar}eA_{\mu}\chi_{\nu},\nonumber\\
F_{\nu\mu}&=&\nabla_{\nu} A_{\mu}-\nabla_{\mu} A_{\nu},\nonumber
\end{eqnarray}
where $\alpha,~A_{\mu},~e~$ and $\nabla_{\mu}$ represents the dimensionless positive parameter,
vector potential, the charge of particle and covariant derivatives, respectively. The non-zero
components of anti-symmetric tensor can be computed as
\begin{eqnarray}
&&\chi^{0}=\frac{-D\chi_{0}+E\chi_{3}}{AD+E^2},~~~\chi^{1}=\frac{1}{B}\chi_{1},
~~~\chi^{2}=\frac{1}{C}\chi_{2},~~~
\chi^{3}=\frac{E\chi_{0}+A\chi_{3}}{AD+E^2},~~\chi^{12}=\frac{1}{BC}\chi_{12},
~\chi^{13}=\frac{1}{BAD+E^2}\chi_{13},\nonumber\\
&&\chi^{01}=\frac{-D\chi_{01}+E\chi_{13}}{B(AD+E^2)},~~~
\chi^{02}=\frac{-D\chi_{02}}{C(AD+E^2)},
~~~\chi^{03}=\frac{(-AD+A^2)\chi_{03}}{(AD+E^2)^2},
~~\chi^{23}=\frac{E\chi_{02}+A\chi_{23}}{C(AD+E^2)},\nonumber
\end{eqnarray}
The WKB approximation is defined as
\begin{equation}
\chi_{\nu}=c_{\nu}\exp\left[\frac{i}{\hbar}\Theta(t, r, \theta, \phi)\right],
\end{equation}
where
\begin{equation}
\Theta(t, r, \theta, \phi)=\Theta_{0}(t, r, \theta,\phi)+{\hbar}\Theta_{1}(t,r,\theta,\phi)+{\hbar}^2\Theta_{2}(t, r, \theta, \phi)+....
\end{equation}
By neglecting the higher order terms and after substituting all the values in Eq. (\ref{xx}), we obtain the set of wave equations as
\begin{eqnarray}
&+&\frac{D}{B(AD+E^2)}\Big[c_{1}(\partial_{0}\Theta_{0})(\partial_{1}\Theta_{0})+\alpha c_{1}
(\partial_{0}\Theta_{0})^{3}(\partial_{1}\Theta_{0})-c_{0}(\partial_{1}\Theta_{0})^{2}
-\alpha c_{0}(\partial_{1}\Theta_{0})^4+c_{1}eA_{0}(\partial_{1}\Theta_{0})\nonumber\\
&+&c_{1}\alpha eA_{0}(\partial_{0}\Theta_{0})^{2}(\partial_{1}\Theta_{0})\Big]
-\frac{E}{B(AD+E^2)}\Big[c_{3}(\partial_{1}\Theta_{0})^2+\alpha c_{3}(\partial_{1}
\Theta_{0})^4-c_{1}(\partial_{1}\Theta_{0})(\partial_{3}\Theta_{0})-\alpha c_{1}
(\partial_{1}\Theta_{0})(\partial_{3}\Theta_{0})^2\Big]\nonumber\\
&+&\frac{D}{C(AD+E^2)}\Big[c_{2}(\partial_{0}\Theta_{0})(\partial_{2}\Theta_{0})
+\alpha c_{2}(\partial_{0}\Theta_{0})^3(\partial_{2}\Theta_{0})-c_{0}(\partial_{2}
\Theta_{0})^2-\alpha c_{0}(\partial_{2}\Theta_{0})^4+c_{2}eA_{0}(\partial_{2}\Theta_{0})\nonumber\\
&+&c_{2}eA_{0}\alpha(\partial_{0}\Theta_{0})^{2}(\partial_{1}\Theta_{0})\Big]
+\frac{AD}{(AD+E^2)^2}\Big[c_{3}(\partial_{0}\Theta_{0})(\partial_{3}\Theta_{0})
+\alpha c_{3}(\partial_{0}\Theta_{0})^{3}(\partial_{3}\Theta_{0})-c_{0}(\partial_{3}\Theta_{0})^{2}\nonumber\\
&-&\alpha c_{0}(\partial_{3}\Theta_{0})^4+c_{3}eA_{0}(\partial_{3}\Theta_{0})+c_{3}eA_{0}
(\partial_{0}\Theta_{0})^{2}(\partial_{3}\Theta_{0})\Big]-m^2\frac{\tilde{D c_{0}}-\tilde{E c_{3}}}{(AD+E^2)}=0,\label{jj}\\
&-&\frac{D}{B(AD+E^2)}\Big[c_{1}(\partial_{0}\Theta_{0})^2+\alpha c_{1}
(\partial_{0}\Theta_{0})^4-c_{0}(\partial_{0}\Theta_{0})(\partial_{1}\Theta_{0})
-\alpha c_{0}(\partial_{0}\Theta_{0})(\partial_{1}\Theta_{0})^{3}
+c_{1}eA_{0}(\partial_{0}\Theta_{0})\nonumber\\
&+&\alpha c_{1}eA_{0}(\partial_{0}\Theta_{0})^3\Big]+\frac{E}{B(AD+E^2)}
\Big[c_{3}(\partial_{0}\Theta_{0})(\partial_{1}\Theta_{0})+\alpha c_{3}
(\partial_{0}\Theta_{0})(\partial_{1}\Theta_{0})^3-c_{1}(\partial_{0}\Theta_{0})(\partial_{3}\Theta_{0})-\alpha c_{1}(\partial_{0}\Theta_{0})(\partial_{3}\Theta_{0})^{3}\Big]\nonumber\\
&+&\frac{1}{BC}\Big[c_{2}(\partial_{1}\Theta_{0})(\partial_{2}\Theta_{0})
+\alpha c_{2}(\partial_{1}\Theta_{0})(\partial_{2}\Theta_{0})^3-c_{1}(\partial_{2}\Theta_{0})^{2}-\alpha c_{1}(\partial_{2}\Theta_{0})^{4}\Big]+\frac{1}{B(AD+E^2)}\Big[c_{3}
(\partial_{1}\Theta_{0})(\partial_{3}\Theta_{0})+\alpha c_{3}\nonumber\\
&\times&(\partial_{1}\Theta_{0})(\partial_{3}\Theta_{0})^3-c_{1}(\partial_{3}\Theta_{0})^2-\alpha c_{1} (\partial_{3}\Theta_{0})^{4}\Big]+\frac{eA_{0}D}{B(AD+E^2)}\Big[c_{1}
(\partial_{0}\Theta_{0})+\alpha c_{1}(\partial_{0}\Theta_{0})^3
-c_{0}(\partial_{1}\Theta_{0})-\alpha c_{0}(\partial_{1}\Theta_{0})^3\nonumber\\
&+&eA_{0}c_{1}+\alpha c_{1}eA_{0}(\partial_{0}\Theta_{0})^{2})\Big]
+\frac{eA_{0}E}{B(AD+E^2)}\Big[c_{3}(\partial_{1}\Theta_{0})+\alpha c_{3}(\partial_{1}\Theta_{0})^3
-c_{1}(\partial_{3}\Theta_{0})-\alpha c_{1}(\partial_{1}\Theta_{0})^3\Big]-\frac{m^2 c_{1}}{B}=0,
\end{eqnarray}
\begin{eqnarray}
&+&\frac{D}{C(AD+E^2)}\Big[c_{2}(\partial_{0}\Theta_{0})^2+\alpha c_{2}
(\partial_{0}\Theta_{0})^{4}-c_{0}(\partial_{0}\Theta_{0})(\partial_{2}\Theta_{0})
-\alpha c_{0}(\partial_{0}\Theta_{0})(\partial_{2}\Theta_{0})^3
+c_{2}eA_{0}(\partial_{0}\Theta_{0})+\alpha c_{2}eA_{0}(\partial_{0}\Theta_{0})^{3}\Big]\nonumber\\
&+&\frac{1}{BC}\Big[c_{2}(\partial_{1}\Theta_{0})^2+\alpha c_{2}
(\partial_{1}\Theta_{0})^{4}-c_{1}(\partial_{1}\Theta_{0})(\partial_{2}\Theta_{0})
-\alpha c_{1}(\partial_{1}\Theta_{0})(\partial_{2}\Theta_{0})^3\Big]-\frac{E}{C(AD
+E^2)}\Big[c_{2}(\partial_{0}\Theta_{0})(\partial_{3}\Theta_{0})\nonumber\\
&+&\alpha c_{2}(\partial_{0}\Theta_{0})^{3}
(\partial_{3}\Theta_{0})-c_{0}(\partial_{0}\Theta_{0})(\partial_{3}\Theta_{0})
-\alpha c_{0}(\partial_{0}\Theta_{0})^3 (\partial_{3}\Theta_{0})+c_{2}eA_{0}(\partial_{3}\Theta_{0})
+\alpha c_{2}eA_{0}(\partial_{3}\Theta_{0})^{3}\Big]\nonumber\\
&+&\frac{A}{C(AD+E^2)}\Big[c_{3}(\partial_{2}\Theta_{0})(\partial_{3}\Theta_{0})+\alpha c_{3}
(\partial_{2}\Theta_{0})^{3}(\partial_{3}\Theta_{0})-c_{2}(\partial_{3}\Theta_{0})^2
-\alpha c_{2}(\partial_{3}\Theta_{0})^4\Big]-\frac{m^2 c_{2}}{C}\nonumber\\
&+&\frac{eA_{0}D}{C(AD+E^2)}\Big[c_{2}(\partial_{0}\Theta_{0})+\alpha c_{2}
(\partial_{0}\Theta_{0})^3-c_{0}(\partial_{2}\Theta_{0})-\alpha c_{0}
(\partial_{2}\Theta_{0})^3+c_{2}eA_{0}+c_{2}\alpha eA_{0}(\partial_{0}\Theta_{0})^2\Big]=0,\\
&+&\frac{(AD)-A^2}{(AD+E^2)^2}\Big[c_{3}(\partial_{0}\Theta_{0})^2+\alpha c_{3}
(\partial_{0}\Theta_{0})^4-c_{0}(\partial_{0}\Theta_{0})(\partial_{3}
\Theta_{0})-\alpha c_{0}(\partial_{0}\Theta_{0})(\partial_{3}\Theta_{0})^{3}
+{eA_{0}c_3}(\partial_{0}\Theta_{0})\nonumber\\
&+&\alpha c_{3}eA_{0}(\partial_{0}\Theta_{0})^{3}\Big]
-\frac{D}{C(AD+E^2)}\Big[c_{3}(\partial_{1}\Theta_{0})^2+\alpha c_{3}
(\partial_{1}\Theta_{0})^{4}-c_{1}(\partial_{1}\Theta_{0})(\partial_{3}\Theta_{0})
-\alpha c_{1}(\partial_{1}\Theta_{0})(\partial_{3}\Theta_{0})^3\Big]\nonumber\\
&-&\frac{E}{C(AD+E^2)}\Big[c_{2}(\partial_{0}\Theta_{0})(\partial_{2}\Theta_{0})+\alpha c_{2}
(\partial_{0}\Theta_{0})^3(\partial_{2}\Theta_{0})-c_{0}(\partial_{2}\Theta_{0})^{2}
+\alpha c_{0}(\partial_{2}\Theta_{0})^4+{eA_{0}c_2}(\partial_{2}\Theta_{0})
+\alpha c_{2}eA_{0}\nonumber\\
&\times&(\partial_{0}\Theta_{0})^{2}(\partial_{2}\Theta_{0})\Big]-\frac{eA_{0}A}{C(AD+E^2)}
\Big[c_{3}(\partial_{2}\Theta_{0})^2+\alpha c_{3}(\partial_{2}
\Theta_{0})^4-c_{2}(\partial_{2}\Theta_{0})(\partial_{3}\Theta_{0})
-\alpha c_{2}(\partial_{0}\Theta_{0})(\partial_{3}\Theta_{0})^{3}\Big]\nonumber\\
&+&\frac{eA_{0}(AD)-A^2}{(AD+E^2)^2}\Big[c_{3}(\partial_{0}\Theta_{0})+\alpha c_{3}(\partial_{0}\Theta_{0})^3
-c_{0}(\partial_{3}\Theta_{0})-\alpha c_{0}(\partial_{3}\Theta_{0})^3+c_{3}eA_{0}
+\alpha eA_{0}(\partial_{0}\Theta_{0})^2\Big]\nonumber\\&-&\frac{m^2 (Ec_{0}-Ac_{3}}{(AD+E^2)}=0.\label{uu}
\end{eqnarray}
Using separation of variables technique, we can choose
\begin{equation}
\Theta_{0}=-\acute{E}t+W(r)+J\phi+\nu(\theta),\label{vv}
\end{equation}
where $\acute{E}=(E-j\omega)$, $E$ denotes the energy of the particle, $J$ represents the particles angular
momentum corresponding to angles $\phi$. After substituting Eq. (\ref{vv}) into
set of wave equations, we get a $4\times4$ matrix
\begin{equation*}
\mathcal{Z}(c_{0},c_{1},c_{2},c_{3})^{T}=0,
\end{equation*}
whose components are given as follows:
\begin{eqnarray}
\mathcal{Z}_{00}&=&\frac{\tilde{-D}}{B(AD+E^2)}\Big[W_{1}^2+\alpha W_{1}^4\Big]
-\frac{D}{C(AD+E^2)}\Big[J^2+\alpha J^4\Big],
-\frac{AD}{(AD+E^2)^2}\Big[\nu_{1}^2+\alpha \nu_{1}^4\Big]-\frac{m^2 D}{(AD+E^2)},\nonumber\\
\mathcal{Z}_{01}&=&\frac{\tilde{-D}}{B(AD+E^2)}\Big[\acute{E}+\alpha
\acute{E}^3+eA_{0}+\alpha eA_{0}\acute{E}^2\Big]W_{1}
+\frac{E}{B(AD+E^2)}+\Big[\nu_{1}+\alpha \nu_{1}^3\Big],\nonumber\\
\mathcal{Z}_{02}&=&\frac{\tilde{-D}}{C(AD+E^2)}\Big[\acute{E}+\alpha
\acute{E}^3-eA_{0}-\alpha eA_{0}\acute{E}^2\Big]J,\nonumber\\
\mathcal{Z}_{03}&=&\frac{\tilde{-E}}{B(AD+E^2)}\Big[W_{1}^2+\alpha W_{1}^4\Big]
-\frac{AD}{C(AD+E^2)^2}\Big[\acute{E}+\alpha \acute{E}^3
-eA_{0}-\alpha eA_{0}\acute{E}^2\Big]\nu_{1}+\frac{m^2E}{(AD+E^2)^2},\nonumber\\
\mathcal{Z}_{11}&=&\frac{\tilde{-D}}{B(AD+E^2)}\Big[\acute{E}^2
+\alpha\acute{E}^4-eA_{0}\acute{E}-\alpha eA_{0}\acute{E}W_{1}^2\Big]
+\frac{E}{B(AD+E^2)}-\frac{m^2}{B}\nonumber\\
&+&\Big[\nu_{1}+\alpha \nu_{1}^3\Big]\acute{E}-\frac{1}{BC}\Big[J^2+\alpha J^4\Big]-
\frac{1}{B(AD+E^2)}\Big[\nu_{1}+\alpha \nu_{1}^3\Big]+\frac{eA_{0}E}{B(AD+E^2)}\Big[\nu_{1}+
\alpha \nu_{1}^3\Big]\nonumber\\
&-&\frac{eA_{0}D}{B(AD+E^2)}\Big[\acute{E}+\alpha \acute{E}^3-eA_{0}
-\alpha eA_{0}\acute{E}^2\Big],~~~~~~\mathcal{Z}_{12}=\frac{1}{BC}[W_{1}+\alpha W_{1}^3]J,\nonumber\\
\mathcal{Z}_{13}&=&\frac{\tilde{-E}}{B(AD+E^2)}\Big[W_{1}+\alpha W_{1}^3\Big]\acute{E}
+\frac{1}{B(AD+E^2)^2}\Big[W_{1}+\alpha W_{1}^3\Big]\nu_{1}
+\frac{EeA_{0}}{B(AD+E^2)}\Big[W_{1}+\alpha W_{1}^3\Big],\nonumber
\end{eqnarray}
\begin{eqnarray}
\mathcal{Z}_{22}&=&\frac{D}{C(AD+E^2)}\Big[\acute{E}^2
+\alpha \acute{E}^4-eA_{0}\acute{E}-\alpha eA_{0}\acute{E}\Big]
-\frac{1}{BC}-\frac{m^2}{C}\nonumber\\
&-&\frac{A}{C(AD+E^2)}\Big[\nu_{1}^2+\alpha \nu_{1}^4\Big]-\frac{eA_{0}D}{C(AD+E^2)}
\Big[\acute{E}+\alpha \acute{E}^3-eA_{0}-\alpha eA_{0}\acute{E}^2\Big]\nonumber\\
&+&\frac{E}{C(AD+E^2)}\Big[\acute{E}
+\alpha \acute{E}^3-eA_{0}-\alpha eA_{0}\acute{E}^2\Big]\nu_{1},\nonumber\\
\mathcal{Z}_{23}&=&\frac{A}{C(AD+E^2)}\Big[J+\alpha J^3\Big]\nu_{1},~~~~~~
\mathcal{Z}_{31}=\frac{1}{B(AD+E^2)}\Big[\nu_{1}+\alpha \nu_{1}^3\Big]W_{1},\nonumber\\
%\mathcal{Z}_{32}&=&\frac{E}{C(AD+E^2)}\Big[J+\alpha J^3\Big]\Acute{E}+
%\frac{A}{C(AD+E^2)}\Big[\nu_{1}+\alpha \nu_{1}^3\Big]J,\nonumber\\
\mathcal{Z}_{33}&=&\frac{(AD-\tilde{A^2})}{(AD+E^2)}\Big[\acute{E}^2
+\alpha \acute{E}^4-eA_{0}\acute{E}-\alpha eA_{0}\acute{E}^3\Big]
-\frac{1}{B(AD+E^2)}\Big[W_{1}^2+\alpha W_{1}^4\Big]\nonumber\\
&-&\frac{A}{C(AD+E^2)}\Big[J^2+\alpha J^4\Big]
-\frac{m^2 A}{(AD+E^2)}-\frac{eA_{0}(AD-\tilde{A^2})}{(AD+E^2)}\Big[\acute{E}
+\alpha \acute{E}^3-eA_{0}\acute{E}^2\Big],\nonumber
\end{eqnarray}
where $J=\partial_{\phi}\Theta_{0}$, $W_{1}=\partial_{r}{\Theta_{0}}$ and $\nu_{1}=\partial_{\theta}{\Theta_{0}}$.
For the non-trivial solution, we set determinant $\mathcal{Z}$ is equal to
zero and get
\begin{eqnarray}\label{a1}
ImW^{\pm}&=&\pm \int\sqrt{\frac{(\acute{E}-eA_{0})^{2}
+X_{1}\Big[1+\alpha\frac{X_{2}}{X_{1}}\Big]}{(AD+E^2)/BD}}dr,\nonumber\\
&=&\pm \pi\frac{(\acute{E}-eA_{0})+\Big[1+\alpha\Xi\Big]}{2\kappa(r_{+})},
\end{eqnarray}
where
\begin{eqnarray}
X_{1}&=&\frac{BE}{(AD+E^2)}\Big[\acute{E}
-eA_{0}\Big]\nu_{1}+\frac{AB}{(AD+E^2)}\nu_{1}^2-Bm^2,\nonumber\\
X_{2}&=&\frac{BD}{(AD+E^2)}\Big[\acute{E}^4-2eA_{0}\acute{E}^3+(eA_{0})^2
\acute{E}^2\Big]-\frac{AB}{(AD+E^2)}\nu_{1}^4-W_{1}^4\nonumber\\
&+&\frac{BE}{C(AD+E^2)}\Big[\acute{E}^3-eA_{0}\acute{E}^2\Big]\nu_{1}.\nonumber
\end{eqnarray}
The tunneling probability for charged vector particles can be given as
\begin{equation}
\Gamma=\frac{\Gamma_{\textmd{emission}}}{\Gamma_{\textmd{absorption}}}=
\exp\left[{-2\pi}\frac{(\acute{E}-eA_{0})}
{\kappa(r_{+})}\right]\Big[1+\alpha\Xi\Big].
\end{equation}
where
\begin{equation}
\kappa(r_{+})=\frac{2\beta^4r^5_{+}+4M\beta r^2_{+}-8r_{+}q^2+a^2(4\beta^4r^3_{+}
-4M\beta)}{2\alpha^2(r^2_+ +a^2)^2}.
\end{equation}
The modified Hawking temperature can be derived after expanding the series
$\Big[1+\alpha\Xi\Big]$ and by using the Boltzmann factor
$\Gamma_{B}=\exp\left[(\acute{E}-eA_{0})/T'_{H}\right]$ as
\begin{equation}
T'_{H}\cong\frac{2\beta^4r^5_{+}+4M\beta r^2_{+}-8r_{+}q^2a^2(4\beta^4r^3_{+}
-4M\beta)}{4\pi \alpha^2(r^2_+ +a^2)^2}\Big[1-\alpha\Xi\Big].
\end{equation}
The modified Hawking temperature of charged black strings depends upon quantum
gravity parameter $\alpha$, mass $M$, charge $q$, spin parameter $a$ and
cosmological constant $\bigwedge$ (i. e., $\beta=-\bigwedge/3$). In the absence
of quantum gravity parameter $\alpha=0$, we observe the temperature of Eq. (\ref{12}).
%The quantum gravity effects reduce the increase in Hawking temperature.

\section{Graphical Analysis}
The section comprises the graphical behavior of modified temperature
of charged black strings in Rastall theory. We examine the effects of quantum
gravity parameters $\alpha$ and spin parameter $a$ (appears due to Rastall
theory) from charged black strings. We also study the physical significance of these plots.
These plots depicts the behavior of $T'_H$ w.r.t horizon $r_+$.

In \textbf{Fig. 2}: (i) indicates the behavior of $T'_H$ for fixed
values of $M=100, \beta=0.01, q=1, a=9, \Xi=10$ and various values of quantum gravity parameter $\alpha$ in the range $0\leq r_{+}\leq8$. One can observe that $T'_H$ goes on decreasing for increasing values of $r_+$.
This is physical phenomenon and represents the stable condition of charged black strings
under the influence of quantum gravity parameter at high temperature.

(ii) shows the behavior of $T'_H$ for fixed $M=100, \beta=0.01, q=8, \alpha=500,
\Xi=10$ and various values of spin parameter $a$. One can see that at first
the $T'_H$ increases very slowly and it reaches at a maximum height with very
high value and the eventually falls down from height and attains an asymptotically
flat form as $T'_H\rightarrow0$ till $r_+\rightarrow\infty$. This is purely stable and
physical form of charged black strings under the effects of spin parameter and quantum
gravity. It is notable that as we increase the values of spin parameter the temperature decreases.
Moreover, the very high $T'_H$ at non-zero $r_+$ indicates the BH remnant.
\begin{center}
\includegraphics[width=8.2cm]{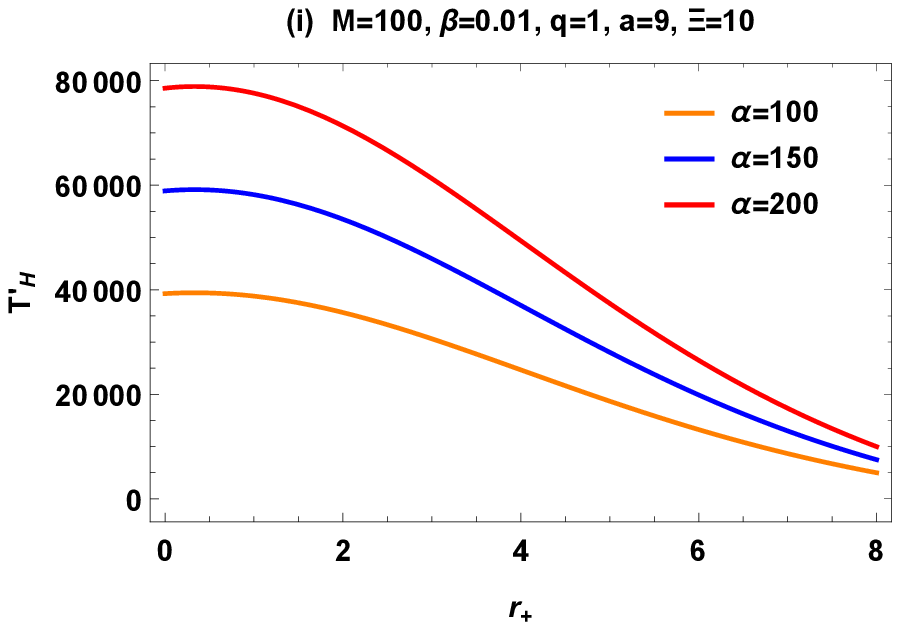}\includegraphics[width=8.2cm]{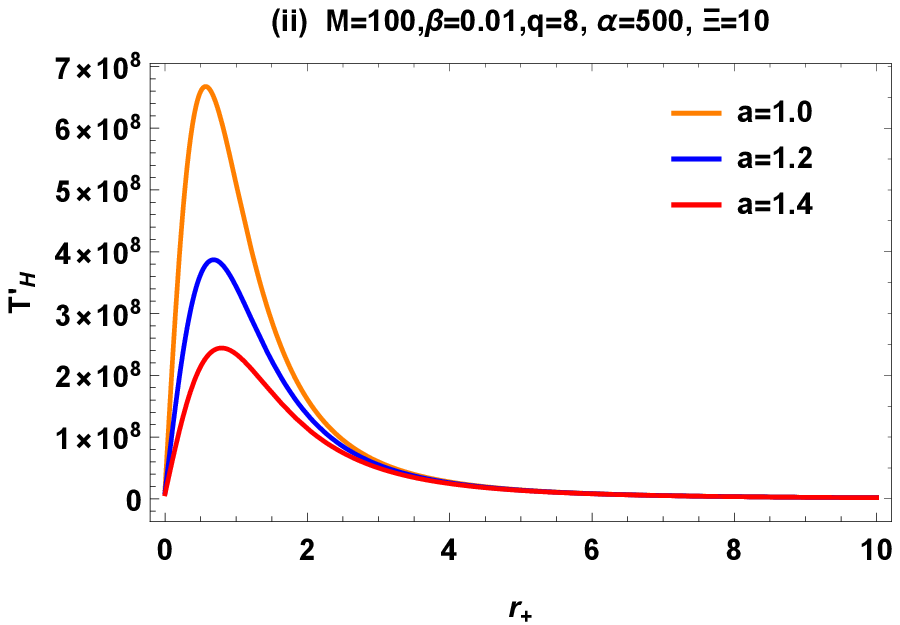}\\
{Figure 2: Hawking temperature $T'_{H}$ versus event horizon $r_{+}$.}
\end{center}

\section{Summary and Discussion}
In our work, we investigated the charged black strings solution in the context of Rastall
theory by applying the Newman-Janis algorithm. After assuming the spin parameter
$(a\rightarrow 0)$ in the Eq. (\ref{ww}), we obtained the black strings solution without Rastall theory in general relativity.
The charged black strings solution in Rastall theory is quite different from the BH solution in general theory of relativity.

The Hawking temperature $T_{H}$ depends on cosmological constant $\bigwedge$,
spin parameter $a$, black string mass $M$ and black
string charge $q$. It is worth mentioning here that for $a=0$, we recovered the Hawking temperature
for charged black strings \cite{21} that is independent of the spin parameter.
It is suggested that the back-reaction affects of the emitted particles on the black string geometry
as well as self-gravitating impacts have been neglected and evaluated Hawking temperature as a term and
yields as black string geometry.
%The particles tunnel through the horizon do not depend on any types of particles.
The Hawking radiation from the charged black strings have different types of particles spins (down, upward or zero spin).
In this procedure, the Hawking temperature is associated to the spin parameter and geometry of charged black strings.
We conclude from the graphical interpretation of temperature $T_{H}$ w.r.t horizon $r_{+}$ that
the charged black strings solution under the influence of Rastall theory for various values
of charge and spin parameter depicts its stable form. Furthermore, we examined the quantum gravity
effects for charged black strings in Rastall theory and derived the modified Hawking temperature.
We also discussed the stable and physical form of charged black strings under the effects of quantum
gravity and spin parameter. The spin parameter which appears due to Rastall theory in charged black
string solution causes the reduction in temperature. The Hawking's phenomenon depicts that with the
emission of more radiations the size of BH radius reduces and we observe BH remnant at very high temperature
with non-zero horizon. We observe this physical phenomenon in all plots which guarantee the stable form of charged black strings.
Since, the conclusion
still holds if background charged black strings geometry is more general.

\end{document}